\documentclass[letterpaper]{article} 
\usepackage{aaai24}  
\usepackage{times}  
\usepackage{helvet}  
\usepackage{courier}  
\usepackage[hyphens]{url}  
\usepackage{graphicx} 
\usepackage{natbib}  
\usepackage{caption} 
\usepackage{soul}
\usepackage[utf8]{inputenc}
\usepackage{amsmath}
\usepackage{amsthm}
\usepackage{amssymb}
\usepackage{algorithm}
\usepackage{algorithmic}
\usepackage{subfigure}
\usepackage{xcolor}
\usepackage{booktabs}
\usepackage[switch]{lineno}
\usepackage{algorithm}
\usepackage{algorithmic}

\usepackage{newfloat}
\usepackage{listings}
\DeclareCaptionStyle{ruled}{labelfont=normalfont,labelsep=colon,strut=off} 
\floatstyle{ruled}
\newfloat{listing}{tb}{lst}{}
\floatname{listing}{Listing}
\title{Jointly Modeling Spatio-Temporal Features of Tactile Signals \\for Action Classification}
\author{
    Jimmy Lin\textsuperscript{\rm 1,2}, Junkai Li\textsuperscript{\rm 1,2}, Jiasi Gao\textsuperscript{\rm 1}, Weizhi Ma\textsuperscript{\rm 1}\footnote{ Corresponding authors.}, Yang Liu\textsuperscript{\rm 1,2}\footnotemark[1]
}
\affiliations{
    \textsuperscript{\rm 1} Institute for AI Industry Research (AIR), Tsinghua University, Beijing, China\\
    \textsuperscript{\rm 2} Department of Computer Science and Technology, Tsinghua University, Beijing, China\\

    jimmy.lin9@gmail.com, li-jk23@mails.tsinghua.edu.cn, bnancy@sina.com,\\ \{mawz, liuyang2011\}@tsinghua.edu.cn
}

\usepackage{bibentry}

\begin{document}

\maketitle

\begin{abstract}
Tactile signals collected by wearable electronics are essential in modeling and understanding human behavior. 
One of the main applications of tactile signals is action classification, especially in healthcare and robotics. 
However, existing tactile classification methods fail to capture the spatial and temporal features of tactile signals simultaneously, which results in sub-optimal performances. 
In this paper, we design \textbf{S}patio-\textbf{T}emporal \textbf{A}ware tactility \textbf{T}ransformer (\textbf{STAT}) to utilize continuous tactile signals for action classification. 
We propose spatial and temporal embeddings along with a new temporal pretraining task in our model, which aims to enhance the transformer in modeling the spatio-temporal features of tactile signals. 
Specially, the designed temporal pretraining task is to differentiate the time order of tubelet inputs to model the temporal properties explicitly. 
Experimental results on a public action classification dataset demonstrate that our model outperforms state-of-the-art methods in all metrics. 
\end{abstract}

\section{Introduction}

Similar to visual and acoustic signals, tactile signals are important for modeling and understanding humans. In recent years, various wearable electronics have been designed to collect tactile signals, which are widely used in multiple scenarios, especially in healthcare and robotics~\cite{zhu2019self,fan2020machine,lou2020highly,DBLP:journals/ral/OkunevichTKT21}.

\begin{figure}[h]
    \centering
    \includegraphics[width=1\linewidth]{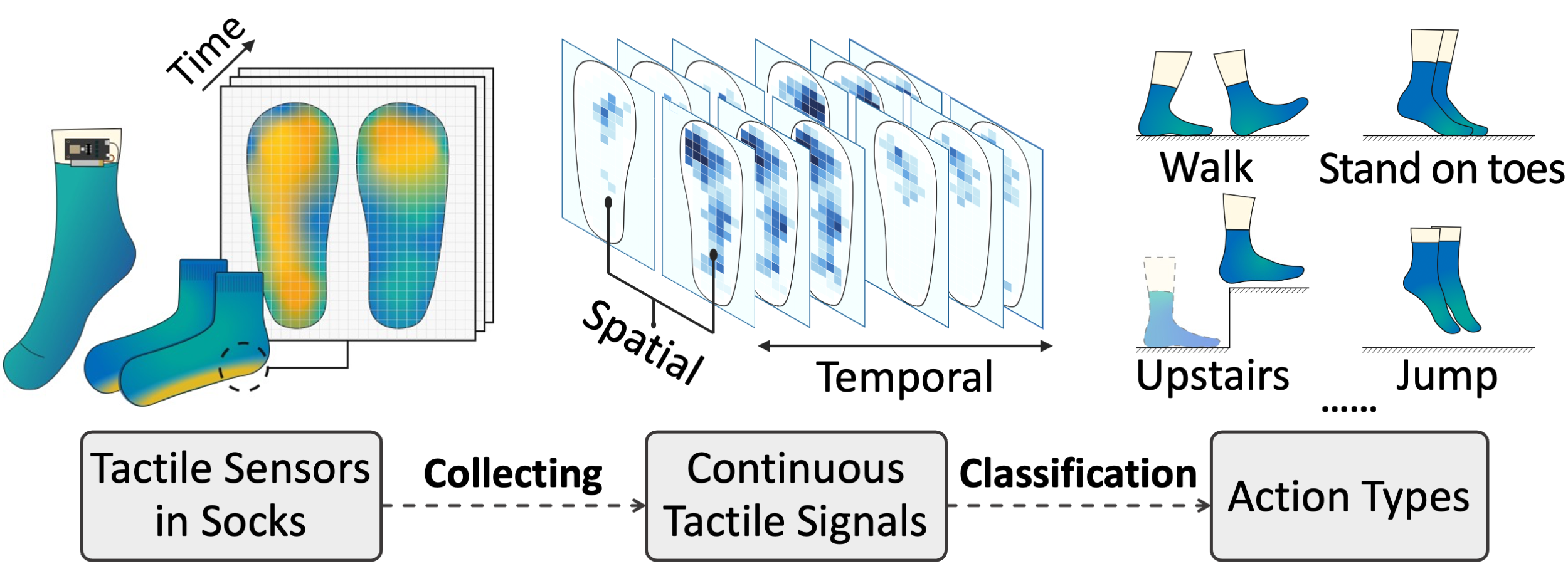}
    \caption{An overview of action classification based on tactile signals collected by wearable electronic socks.}
    \label{fig:signal overview}
\end{figure}

The collected tactile signals can be utilized for different purposes, and one of their main applications is the action classification task.  
Sundaram et al.~\shortcite{DBLP:journals/nature/SundaramKLZ0M19} propose to identify hand actions by tactile signals with sensors in gloves. 
Luo et al.~\shortcite{luo2021learning} and Wicaksono et al.~\shortcite{wicaksono20223dknits} use wearable electronic socks to collect tactile signals for feet action classification.
Figure~\ref{fig:signal overview} is an example, where the continuous tactile signals are collected by e-textile sensors in socks, and then used to classify the action (e.g., walking, etc.).

Tactile signals are spatially and temporally sensitive, hence utilizing their spatio-temporal features is important for action classification.
Firstly, tactile signals are spatially sensitive as they are not \textit{translation invariant}. The same signals in different positions (i.e., collected by various sensors) indicate distinct actions. 
For example, the same signals collected by sensors located in different positions should be classified as standing on toes or heels, respectively. 
Secondly, tactile signals are temporally sensitive as they are collected regularly with high frequency, e.g., in 10HZ (10 data points per second), and the time order of these signals is informative. For example, if ignoring the order of collected signals, two signal sequences collected by the same sensor from distinct actions may be seen as identical actions (i.e., same elements with different orders), which becomes useless in classification. 

\begin{figure*}[h]
    \centering
    \includegraphics[width=0.78\linewidth]{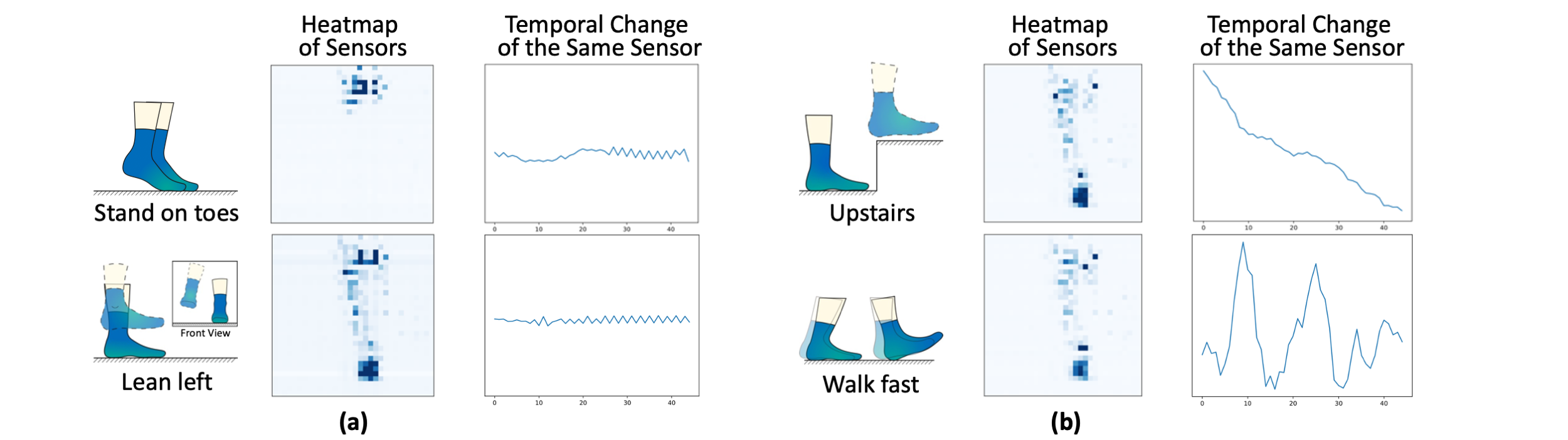}
    \caption{Empirical study of actions in a tactile dataset. Heatmaps are the averaged results of all samples collected by sensors in the left foot, and the tactile sensor of Figure 2(a) and 2(b) is located at positions (5,20) and (28,19) of the left foot, respectively. }
    \label{fig:compare}
\end{figure*}

Furthermore, we want to point out that jointly modeling spatial and temporal features is essential for tactile signals in action classifications. 
We conduct an empirical study on a real-scenario dataset~\cite{luo2021learning}, and
draw the spatial and temporal features of different actions in Figure~\ref{fig:compare}. 
The heatmap of each action shows the averaged results of all samples, which indicates spatial features. The temporal change of a specific sensor shows the averaged sequence data of all samples collected by this sensor, which indicates temporal features.
As shown in Figure \ref{fig:compare}(a), two actions, stand on toes and lean left, have similar temporal features but different spatial features. However, in Figure \ref{fig:compare}(b), two actions, upstairs and walk fast, have similar spatial features but different temporal features. 
These observations verify tactile signals' spatio-temporal features, and further indicate that only using one of them is inadequate for classification.

However, existing tactile methods lack the ability to capture the mentioned temporal and spatial nature of tactile signals simultaneously. 
On the one hand, most previous tactile-related studies adopt CNN-based methods to model the tactile signal frames and then combine them by concatenating or sequential models, which fail to jointly capture their translation variance and temporal properties~\cite{luo2021learning,DBLP:journals/nature/SundaramKLZ0M19,DBLP:conf/iros/GaoTL020}. 
On the other hand, various transformer models have been designed to handle different continuous signals. But most of them~\cite{DBLP:conf/kdd/ZerveasJPBE21,DBLP:journals/corr/abs-2203-12602,amiridi2022latent} focus on temporal features, which are inadequate to model tactile signals' spatial nature, especially the \textit{translation variant} property.

In this paper, we design a \textbf{S}patio-\textbf{T}emporal \textbf{A}ware tactility \textbf{T}ransformer (\textbf{STAT}) to utilize tactile signals for action classification, which utilizes their temporal and spatial features simultaneously.
We design spatial and temporal embeddings to explicitly model the translation variant and sequential features of tactile signals, respectively. Additionally, we introduce a temporal pretraining task to enhance the modeling of temporal features by distinguishing the time order of signal tubelets.  
After pretraining the STAT transformer, the embedding of the [CLS] token is utilized for action classification. 
Experimental results on tactile show that our model outperforms all baseline methods in all metrics, including state-of-the-art multivariate and video classification models. 
Further analyses verify the effectiveness of the proposed pretraining task and embeddings.
To the best of our knowledge, this is the first transformer model designed for tactile signals by jointly modeling spatio-temporal features, which can be applied to various tactile-related scenarios.

\section{Related Work}
\subsection{Action Classification with Tactile Signals}
In recent years, various wearable electronics have been designed to model user actions based on tactile signals in different scenarios. 
Luo et al.~\shortcite{luo2021learning} design wearable electronic socks to classify user walking actions. 
Noh et al.~\shortcite{noh2021xgboost} use tactile signals in healthcare scenarios, which predict the fall risk of users. 
Robotic studies also point out that modeling tactile signals are important in understanding humans ~\cite{DBLP:conf/ijcai/KragicGKJ018,DBLP:journals/scirobotics/NegreJVCKB18}. 

Despite the importance of tactile signals in various scenarios, we find that previous tactile classification models are unable to capture the spatial and temporal properties of tactile signals simultaneously.
Sundaram et al.~\shortcite{DBLP:journals/nature/SundaramKLZ0M19} use CNN to capture the embedding of each frame and simply concatenate them for action classification. 
Recent studies enhance this method by adopting a GRU/LSTM model rather than concatenation to model the sequential features~\cite{luo2021learning,DBLP:journals/ral/OkunevichTKT21,DBLP:conf/iros/GaoTL020}. 
Cao et al.~\shortcite{cao2020spatio} introduces temporal attention operation combined with spatial features in separate phases.
However, as CNNs are designed to utilize the translation invariance features, they fail to capture tactile signals' translation variance. 

Different from previous studies, we design a new transformer model to jointly capture the spatio-temporal features of tactile signals for action classification.

\subsection{Transformers for Continuous Signals}
Transformer models~\cite{vaswani2017attention} have achieved great success in continuous signal classification tasks, e.g., videos and multivariate continuous signals~\cite{DBLP:journals/corr/abs-2203-12602,DBLP:conf/kdd/ZerveasJPBE21,zhao2022alignment}. We briefly review related transformers here, especially video transformers, as the input shape of videos is similar to tactile signals.

Existing transformer models fail to utilize the spatial and temporal features of tactile signals simultaneously.
On the one hand, most video transformers use the visual transformer~\cite{DBLP:conf/iclr/DosovitskiyB0WZ21} as a backbone model, and further propose new masking or input strategies~\cite{DBLP:conf/iccv/Arnab0H0LS21,DBLP:conf/icml/BertasiusWT21,51457,DBLP:conf/cvpr/LiuFC0GGW22,DBLP:conf/cvpr/YanXALZ0S22}. Recent models propose to capture the spatio-temporal features of videos, e.g., VideoMAE~\cite{DBLP:journals/corr/abs-2203-12602}, SSTSA~\cite{9834306}. 
However, as video transformers aim to model the translation invariant of videos, they only use position embeddings in encoding, which fail to model the translation variant spatial property of tactile signals. 
On the other hand, most transformer methods proposed for multivariate continuous signals focus on modeling their temporal features, while ignoring the spatial relations among different signals (i.e., where the signals are collected), such as TST~\cite{DBLP:conf/kdd/ZerveasJPBE21} and the transformer proposed by Hannan et al~\shortcite{hannan2021deep}.
Furthermore, most transformer models rely on the masking and reconstruction pretraining task~\cite{DBLP:conf/naacl/DevlinCLT19,DBLP:conf/iclr/Bao0PW22,DBLP:journals/corr/abs-2203-12602}, which cannot explicitly capture the temporal/spatial features of continuous signals. 

Although these transformers are not designed for tactile signals, we will use them to verify the effectiveness of STAT.

\begin{figure*}[t]
    \centering
    \includegraphics[width=0.87\textwidth]{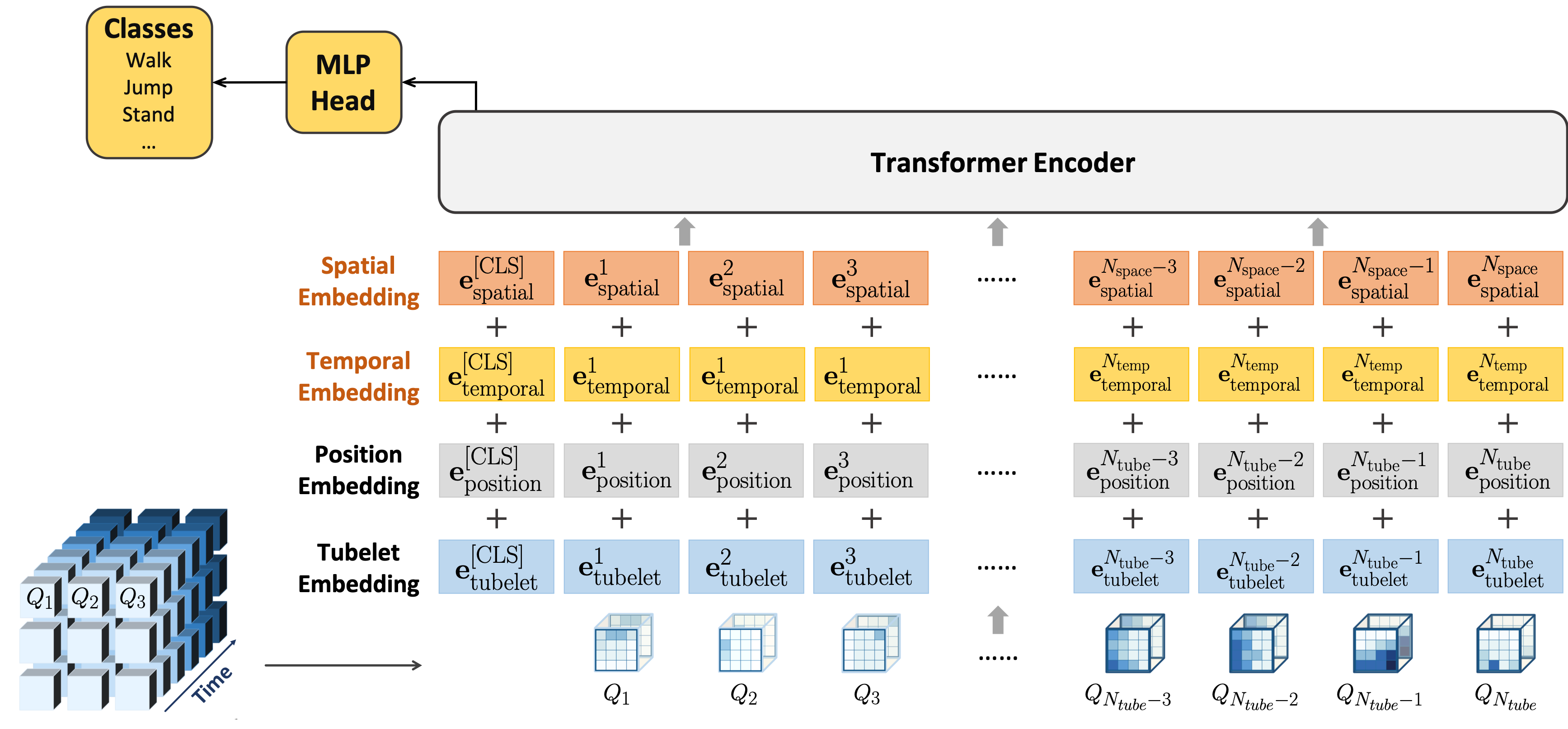}
    \caption{An overview of STAT model. Spatial and temporal embeddings are designed to jointly capture both properties. }
    \label{fig:model overview}
\end{figure*}

\section{Approach}
\subsection{Problem Statement}
Our goal is to utilize tactile signals to classify user actions, where the data can be collected by various wearable electronics. 
The wearable devices often arrange sensors as a matrix, so we define the data matrix collected in a specific time point as a frame. Then, the task is defined as follows.

Given a tactile signal tensor $\mathcal{X} \in \mathbb{R}^{ C \times T \times H \times W }$, where $C$ represents the number of wearable devices, $T$ represents the length of signal sequences (i.e., $T$ frames), $H$ and $W$ mean the number of sensors in each column/row (i.e., the shape of frames), respectively.
An example of tactile signals is shown in Figure \ref{fig:cube data}(a). 
$\mathcal{X}_{c_i, t_j, h_k, w_l}$ represents the value collected by the sensor of device $c_i$ in position $(h_k, w_l)$ at time $t_j$. 
Each tactile segment $\mathcal{X}$ has an activity label $y$, and the total number of activity types is $M$.
Our target is to accurately classify the given tactile signal $\mathcal{X}$ to its label $y$.

\subsection{Overview}
We propose a spatio-temporal aware tactility transformer for the action classification task based on tactile signals, which is named STAT.
A new pretraining task and two extra embeddings are designed to capture the temporal and spatial features of tactile signals jointly in STAT.

Firstly, the designed spatio-temporal aware transformer encoder is introduced. We convert the tactile signal tensor to a tubelet sequence. Besides the widely used tubelet and position embeddings, we propose to add spatial and temporal embeddings to capture each tubelet's temporal and spatial features, respectively. Then, multi-layer transformer encoders are adopted to calculate the representations of tactile signals.
Then, the adopted pretraining tasks are defined. Aside from the common masking and reconstruction loss, we designed a temporal pretraining task to explicitly discriminate the time order of tubelet pairs. 
Finally, we show how to adopt our model for action classifications.

\subsection{Spatio-Temporal Aware Transformer Encoder}
We will introduce the designed spatio-temporal aware transformer encoder shown in Figure~\ref{fig:model overview}. 
To simplify the notations, we only show the process for handling tactile tensor collected from one wearable device (i.e., $\mathcal{X} \in \mathbb{R}^{T \times H \times W }$), as we can easily expand our model to $C$-channel transformers to utilize signals collected by $C$ devices.

\subsubsection{Tubelet Inputs}
\begin{figure}[h]
    \centering
    \includegraphics[width=1\linewidth]{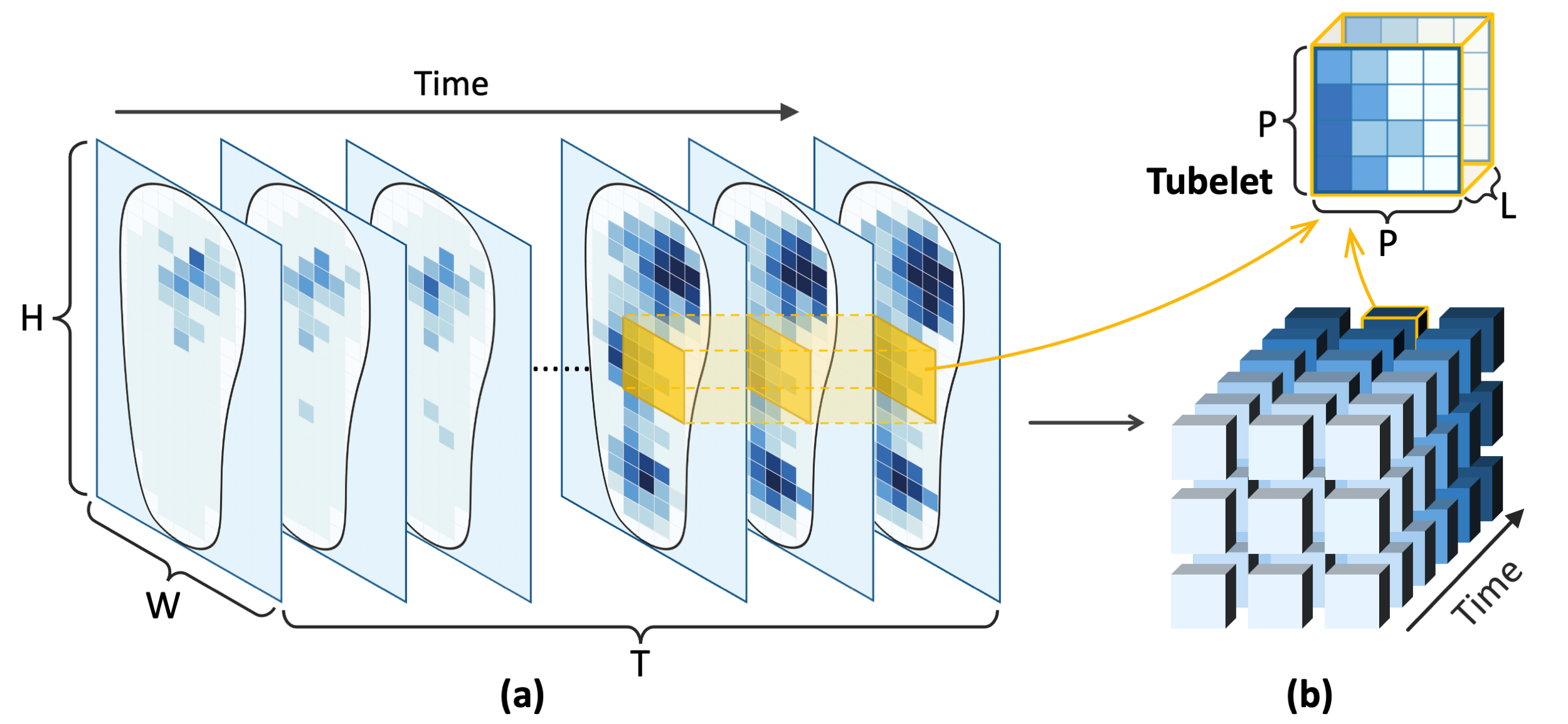}
    \caption{(a) Visualization of tactile signal $\mathcal{X} \in \mathbb{R}^{T \times H \times W }$. (b) Tubelet inputs, where each tubelet $\mathcal{Q} \in \mathbb{R}^{L \times P \times P}$.}
    \label{fig:cube data}
\end{figure}

As the spatial and temporal dimensions of the tactile signals can be redundant, directly adopting the whole data in classification may result in reduced efficiency.
Motivated by previous video transformer models that convert the video clip into tubelets to alleviate the spatio-temporal redundancy, we follow these studies by transferring the tactile signals into a tubelet sequence~\cite{https://doi.org/10.48550/arxiv.2103.15691,https://doi.org/10.48550/arxiv.2106.13230,https://doi.org/10.48550/arxiv.2104.11227,tu2022video}. 
We define a tubelet as $\mathcal{Q} \in \mathbb{R}^{L \times P \times P}$, where $L$ represents its sequence length (i.e., the number of frames) and $P$ represents the patch size (i.e., height and width). Figure~\ref{fig:cube data}(b) shows some examples of the converted tubelets, and the total number of tubelets for a tactile signal tensor $\mathcal{X}$ is $N_{\mathrm{tube}}=THW/(LP^2)$.

\subsubsection{Spatio-Temporal Enhanced Tubelet Embeddings}
Most video transformer models adopt the tubelet embeddings and position embeddings as the input of transformer encoders~\cite{DBLP:conf/iclr/DosovitskiyB0WZ21,DBLP:journals/corr/abs-2203-12602}. However, due to the fact that tactile signals do not have the translation invariance property as images/videos, simply adopting these settings cannot capture the spatial features of tactile signals.
Additionally, jointly modeling spatial and temporal features are also essential in distinguishing actions, as shown in Figure~\ref{fig:compare}.
Thus, we propose to add spatial embeddings and temporal embeddings for each tubelet to capture the spatio-temporal features of tactile signals jointly.

\noindent\textbf{Spatial Embeddings.}
Each tubelet is collected from a patch of sensors, and the sensors are located in certain positions. We use a spatial embedding $\mathbf{e}^{\mathrm{spatial}}_k$ to represent where the tubelet signal is collected from, so that the spatial features will be encoded to explicitly model the translation variance. Tubelets collected by the same batch of sensors will get the same spatial embeddings, and the number of spatial embedding types is $N_{\mathrm{space}} = HW/P^2$.

Following the traditional calculation of position embeddings~\cite{vaswani2017attention}, we utilize the sinusoidal positional encoding table to calculate the spatial embedding ${\mathbf{e}^{\mathrm{spatial}}_k}$, where $k$ represents the spatial position and $k \in \{1,2,..,N_{\mathrm{space}}\}$. 
The calculation is defined in Equation~\eqref{sinusoidal table}:
\begin{equation} \label{sinusoidal table}
\begin{split}
    {\mathbf{e}_{(k,2d)}^{\mathrm{spatial}}} = \sin(\frac{k}{10000^{\frac{2d}{D}}}) \\
    {\mathbf{e}_{(k,2d+1)}^{\mathrm{spatial}}} = \cos(\frac{k}{10000^{\frac{2d}{D}}})
\end{split}
\end{equation}
Where $D$ represents the embedding dimensions, and $\mathbf{e}^{\mathrm{spatial}}_{(k,d)}$ refers to the $d$-th dimension of  $\mathbf{e}^{\mathrm{spatial}}_{k}$ ($d \in \{0,1,2,3...,D\}$).
Through this encoding process, the spatial embeddings can provide the transformer encoder with spatial knowledge of the tactile signals, which contributes to modeling the translation variant features.

\noindent\textbf{Temporal Embeddings.} 
For tactile signal tubelets, their temporal features are important in distinguishing various actions. We propose temporal embeddings to represent the location of the tubelet in the time sequence, which refers to when the tubelet is collected. 
 
Similar to the spatial embeddings, we use the sinusoidal positional encoding Equation~\eqref{sinusoidal table} to generate the temporal embedding $\mathbf{e}^{\mathrm{temporal}}_{k}$. 
The number of temporal embedding types is $N_{\mathrm{temp}} = T/L$, and tubelets collected in the same frames have the same $\mathbf{e}^{\mathrm{temporal}}_{k}$, where $k \in \{1,2,...,N_{\mathrm{temp}}\}$. 

These new embeddings are used to enhance the model with additional information about the spatial and temporal properties of tactile signals jointly, and Figure~\ref{fig:model overview} shows an example.
Then, we aggregated the proposed two embeddings with tubelet and position embeddings to calculate the input matrix $\mathbf{E}^\mathrm{input}$ of transformer encoders by Equation~\eqref{Embedding}. Ultimately, the aggregation of embeddings allows for the simultaneous embedding of spatial and temporal properties.
\begin{equation} \label{Embedding}
\begin{split}
    \mathbf{E}^\mathrm{input} = \mathbf{E}^\mathrm{tubelet} + \mathbf{E}^\mathrm{position} + \mathbf{E}^\mathrm{spatial} + \mathbf{E}^\mathrm{temporal}
\end{split}
\end{equation}

As shown in Figure~\ref{fig:model overview}, we append a [CLS] token at the beginning of the tubelet sequence, which is often used to represent the whole embedding sequence in transformer models. The tubelet, position, temporal, and spatial embeddings of this token are randomly initialized and optimized during training. Hence, we have $\mathbf{E}^{\mathrm{input}} = [\mathbf{E}^{\mathrm{input}}_{[\mathrm{CLS}]},\mathbf{E}^{\mathrm{input}}_{\mathcal{Q}_1},...,\mathbf{E}^{\mathrm{input}}_{\mathcal{Q}_{N_{\mathrm{tube}}}}] \in \mathbb{R}^{(N_{\mathrm{tube}}+1) \times D}$.

\subsubsection{Transformer Encoders}
We utilize the classical transformer encoder \cite{vaswani2017attention} as the backbone network, whose effectiveness has been verified in various domains and tasks~\cite{DBLP:conf/naacl/DevlinCLT19,https://doi.org/10.48550/arxiv.2103.15691,DBLP:conf/iclr/DosovitskiyB0WZ21}.

Our transformer encoder takes in $\mathbf{E}^{\mathrm{input}}$ defined in the previous subsection. 
As the transformer encoder often consists of $K$ transformer layers, we note the primary input $\mathbf{E}^{\mathrm{input}}$ as $\mathbf{E}^{(0)}$,  and $\mathbf{E}^{(k)} = \mathrm{Transformer_k}(\mathbf{E}^{(k-1)})$, where $k \in \{1,2,...,K\}$. The output of the final layer $\mathrm{Transformer_K}$ is the encoded representation of input tokens, and $\mathbf{E}^{(K)}_{[\mathrm{CLS}]}$ is the final embedding of tactile signals.

\subsection{Pretraining Tasks}

\begin{figure}[h]
    \centering
    \includegraphics[width=0.9\linewidth]{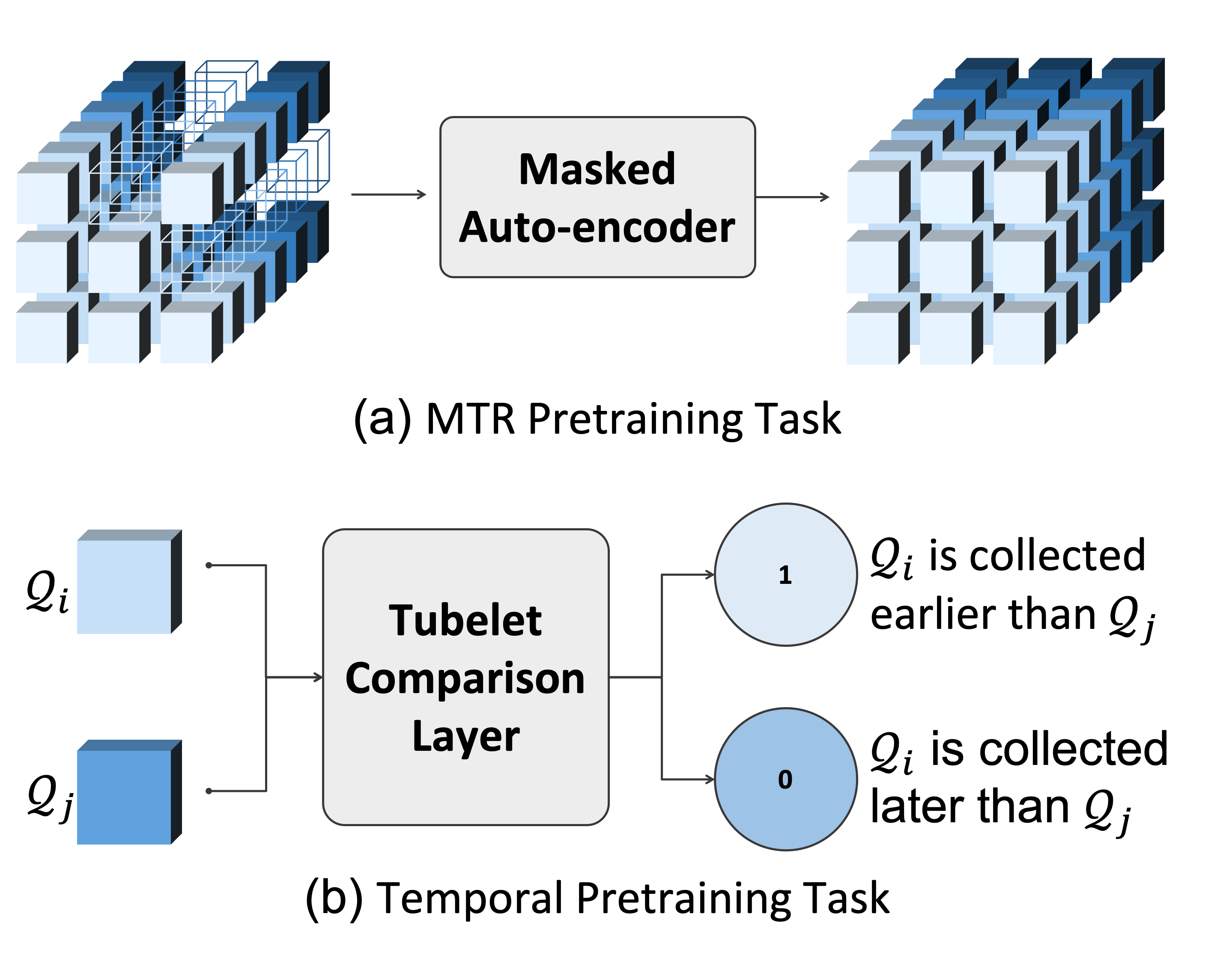}
    \caption{Illustrations of the adopted two pretraining tasks.}
    \label{fig:pretraining}
\end{figure}

Pretraining has been verified to be an effective technique to enhance transformer models in various scenarios, e.g., BERT for text~\cite{DBLP:conf/naacl/DevlinCLT19}, BEIT for image~\cite{DBLP:conf/iclr/Bao0PW22}, and VideoMAE for video~\cite{DBLP:journals/corr/abs-2203-12602}.  
To achieve better classification performances, we choose to pretrain our STAT model before applying it to action classifications.
We propose to use two pretraining tasks here, as shown in Figure~\ref{fig:pretraining}. The first one is the masked tubelet reconstruction (MTR) task, which aims to reconstruct the masked input tubelets, which is also used in previous video transformers~\cite{DBLP:journals/corr/abs-2203-12602}.
The other one is our designed temporal pretraining task to explicitly model the temporal features of tactile signal tubelets. Although temporal embeddings are helpful in capturing temporal properties, we prefer to add a specific pretraining task due to the importance of temporal features in distinguishing different actions.

\subsubsection{Masked Tubelet Reconstruction}
As shown in Figure~\ref{fig:pretraining}(a), we use a masked auto-encoder to reconstruct the input signals, which is the MTR task in previous studies~\cite{DBLP:journals/corr/abs-2203-12602,DBLP:conf/iccv/Arnab0H0LS21}.
MTR randomly masks tubelets from videos and reconstructs them in pretraining, and its loss function is defined as follows:
\begin{equation} \label{MTR}
    \mathcal{L}_\mathrm{MTR} = \frac{1}{|\mathbb{M}|}\sum_{t \in \mathbb{M}} |V_t - \hat{V}_t|^2
\end{equation}
Where $\mathbb{M}$ is the set of masked tubelets' indexes, $V$ is the input video, and $\hat{V}$ is the reconstructed one.
Specially, we adopt spatial-based random masking instead of randomly masking all tubelets. In this strategy, we randomly select sensor groups from the $N_\textrm{space}$ types for masking.
All signals collected by the chosen masking sensors (i.e., tubelets with the same spatial embeddings) will be masked.
The motivation is that this masking strategy will contribute to better utilizing the spatial features among sensors. 
For the mask ratio, we leave it as a hyper-parameter study. 

\subsubsection{Temporal Pretraining}
Our self-supervised temporal pretraining task enhances transformer encoders by training to distinguish the time order of two randomly selected tubelets, so that the temporal features can be maintained in the model, as shown in Figure \ref{fig:pretraining}(b).

Firstly, two tubelets $\mathcal{Q}_i$ and $\mathcal{Q}_j$ are randomly selected from the whole set. Note that  
we should make sure $\mathcal{Q}_i$ and $\mathcal{Q}_j$ are collected at different times, so the temporal embeddings of them are different (i.e., $\mathbf{e}_{\mathcal{Q}_i}^{\mathrm{temporal}} \neq \mathbf{e}_{\mathcal{Q}_j}^{\mathrm{temporal}}$).
Then, we use the encoded embeddings $\mathbf{E}^{(K)}_{\mathcal{Q}_i}$ and $\mathbf{E}^{(K)}_{\mathcal{Q}_j}$ of tubelet $\mathcal{Q}_i$ and $\mathcal{Q}_j$ to identify the time order of them. If tubelet $\mathcal{Q}_i$ is collected earlier than 
$\mathcal{Q}_j$, the identification result should be $y^{\mathrm{temp}}_{i,j} = 1$, otherwise 0.

We choose a simple but effective way to optimize this task. We concatenate the two embeddings $\mathbf{E}^{(K)}_{\mathcal{Q}_i}$ and $\mathbf{E}^{(K)}_{\mathcal{Q}_j}$, and use a linear layer with a sigmoid activation function to predict the time order $\hat{y}^{\mathrm{temp}}_{i,j}$.  A binary cross-entropy loss is utilized to optimize this pretraining task. 
Moreover, for each tactile signal tensor, only randomly selecting one pair of tubelets for pretraining is not enough. So $N_{\mathrm{comp}}$ tubelet pairs are randomly selected and used in pretraining, i.e., ($\mathcal{Q}_{i_1}, \mathcal{Q}_{j_1}$), ..., ($\mathcal{Q}_{i_{N_{\mathrm{comp}}}}$, $\mathcal{Q}_{j_{N_{\mathrm{comp}}}}$). 
Finally, the loss function is formally defined in Equation~\eqref{Frame Comparison Pretraining}.
\begin{equation} \label{Frame Comparison Pretraining}
\small
\centering
\begin{split}
    \hat{y}_{i_n,j_n}^{\mathrm{temp}} =& \sigma(\textbf{W}_{\mathrm{frame}}(\mathbf{E}^{(K)}_{\mathcal{Q}_{i_n}} \oplus \mathbf{E}^{(K)}_{\mathcal{Q}_{j_n}})^\top \\
    \mathcal{L}_{\mathrm{temp}} =& -(y_{i_n,j_n}^{\mathrm{temp}}\log(\hat{y}_{i_n,j_n}^{\mathrm{temp}}) \\
    & + (1 - y_{i_n,j_n}^{\mathrm{temp}})\log(1 - \hat{y}_{i_n,j_n}^{\mathrm{temp}}))
\end{split}
\end{equation}
Where $\oplus$ means vector concatenation, $\sigma$ is a sigmoid activation function, $n \in \{1,2,...,N_{\mathrm{comp}}\}$ and $\textbf{W}_{\mathrm{frame}} \in \mathbb{R}^{1\times2D}$.

To simultaneously utilize these two tasks in pretraining, we add an extra setting that only randomly selects unmasked tubelets, so we can optimize them together.
Specifically, we aggregate the MTR loss with our temporal loss through a weight coefficient $\beta$, which is a hyper-parameter. The final pretraining loss is defined as follows:
\begin{equation} \label{loss}
    \mathcal{L}_\mathrm{pretrain} = \mathcal{L}_{\mathrm{MTR}} + \beta \mathcal{L}_{\mathrm{temp}}
\end{equation}

With the spatial-based random mask strategy in the MTR task and the designed temporal pretraining task, we enhance the representation ability of transformer encoders to better capture the spatial and temporal properties of tactile signals jointly.
Similar to the pretraining of video transformers, we only use the training set of tactile signals for pretraining, as there lacks large scale open tactile datasets.

\subsection{Fine-Tuning for Action Classification}
After introducing our STAT model and pretraining tasks, we will show how to train STAT for action classification.

We follow the approach of other transformer models by using the embedding of  [CLS] token to represent the entire signal sequence. Firstly, we take the embedding of [CLS] token from the last transformer layer block (i.e., $\mathbf{E}^{(K)}_{[\mathrm{CLS}]}$), which represents the whole input signal. Then, we add a linear layer on the top of this embedding to classify it into action types (shown in Figure~\ref{fig:signal overview}). The loss function is:
\begin{equation}
\begin{split}
    \hat{\mathbf{y}}_{i} =& \delta(\textbf{W}_\mathrm{c}(\mathbf{E}^{(K)}_{[\mathrm{CLS}]})^\top + \textbf{b}_\mathrm{c}) \\
    \mathcal{L} =& \mathrm{CrossEntropy}(\hat{\mathbf{y}}_{i}, \mathbf{y}_{i})
\end{split}
\end{equation}
Where $\delta$ is the softmax activation function, $\textbf{W}_{\mathrm{c}} \in \mathbb{R}^{M\times2D}$ and $\textbf{b}_\mathrm{c} \in \mathbb{R}^{1\times M}$, and $\mathbf{y}_{i}$ is a one-hot vector where only the index of the true label is 1.

\begin{table}[]
\centering
\small
\begin{tabular}{lr|lr}
\toprule
\textbf{Actions} & \textbf{\#Samples} & \textbf{Actions} & \textbf{\#Samples} \\ \midrule
Downstairs       & 4,942             & Stand\_toes      & 3,978             \\
Jump             & 3,090             & Upstairs         & 5,025             \\
Lean\_left       & 5,047             & Walk             & 6,078             \\
Lean\_right      & 5,011             & Walk\_fast       & 5,360              \\  
 
Stand            & 5,024             &    \\ \bottomrule
\end{tabular}%
\caption{Statistics of each type of action.}
\label{dataset}
\end{table}

\begin{table*}[h]
    \centering
    \small
    \begin{tabular}{lrrr}
      \toprule
         \textbf{Models} & \textbf{ACC@1} & \textbf{ACC@3} & \textbf{Macro-F1}   \\ 
      \midrule

        CNN\&GRU~\cite{luo2021learning} &  \underline{0.8794}\tiny{$\pm$0.0280} & 0.9497\tiny{$\pm$0.0183} & \underline{0.8743}\tiny{$\pm$0.0319}  \\ \addlinespace
        TST~\cite{DBLP:conf/kdd/ZerveasJPBE21} &  0.8701\tiny{$\pm$0.0252} & \underline{0.9637}\tiny{$\pm$0.0147} & 0.8660\tiny{$\pm$0.0272}  \\ \addlinespace
         VideoMAE~\cite{DBLP:journals/corr/abs-2203-12602} &  0.7705\tiny{$\pm$0.0906} & 0.9287\tiny{$\pm$0.0177} & 0.7521\tiny{$\pm$0.1027}  \\ \addlinespace
        
        \midrule

         STAT w/o pretraining &  0.8050\tiny{$\pm$0.0549} & 0.9528\tiny{$\pm$0.0225} & 0.7946\tiny{$\pm$0.0652} \\ \addlinespace
         STAT & \textbf{0.9033}\tiny{$\pm$0.0098} & \textbf{0.9830}\tiny{$\pm$0.0081} & \textbf{0.9015}\tiny{$\pm$0.0104} \\ \addlinespace
      \bottomrule
    \end{tabular}
        \caption{{Overall performances of all models.}}
\label{main_result}
\end{table*}

\begin{figure*}[]
    \centering
    \subfigure[CNN\&GRU]{
        \includegraphics[width=4cm]{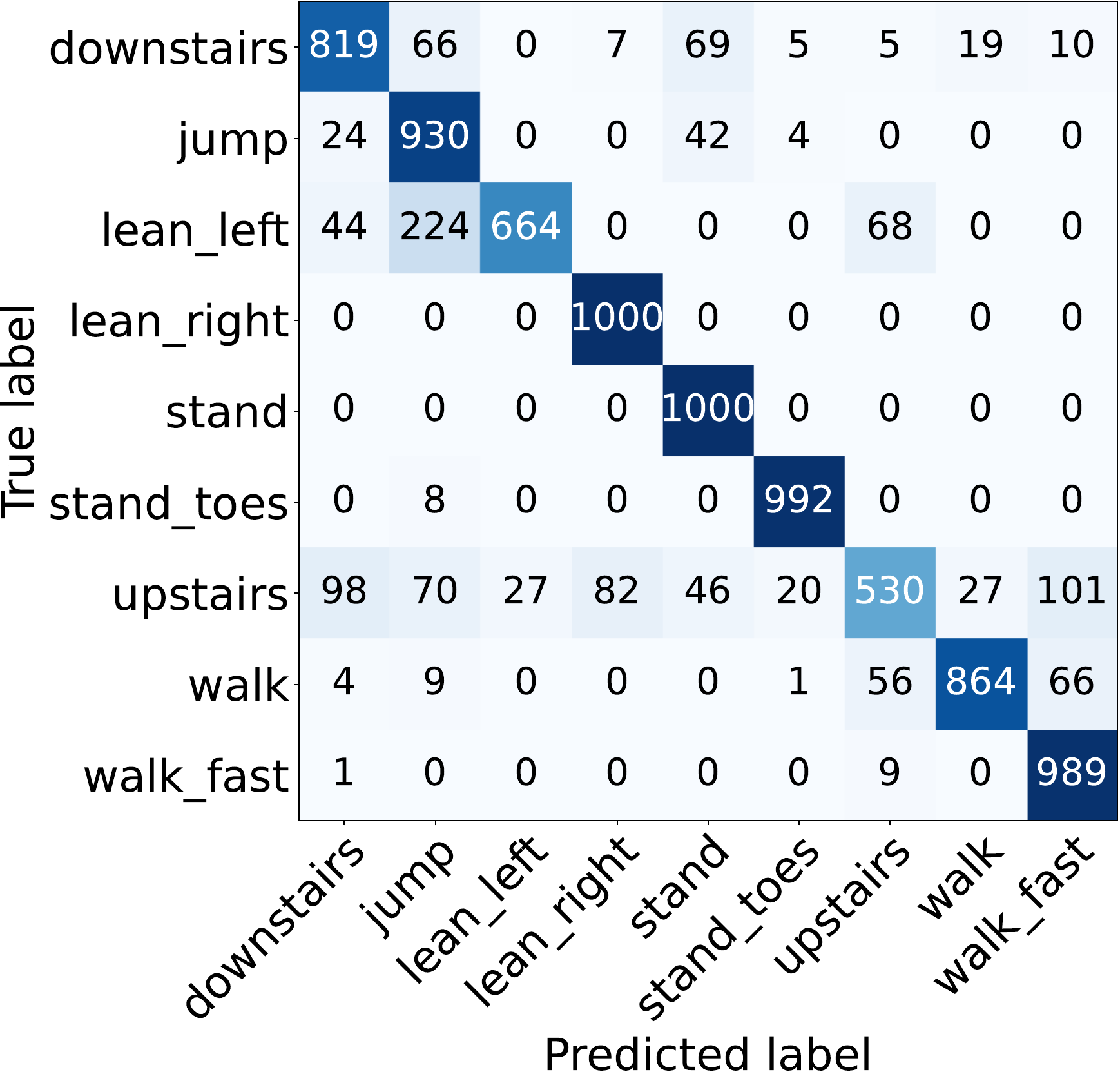} 
        \label{fig:CNN}
    }
    \subfigure[TST]{
        \includegraphics[width=4cm]{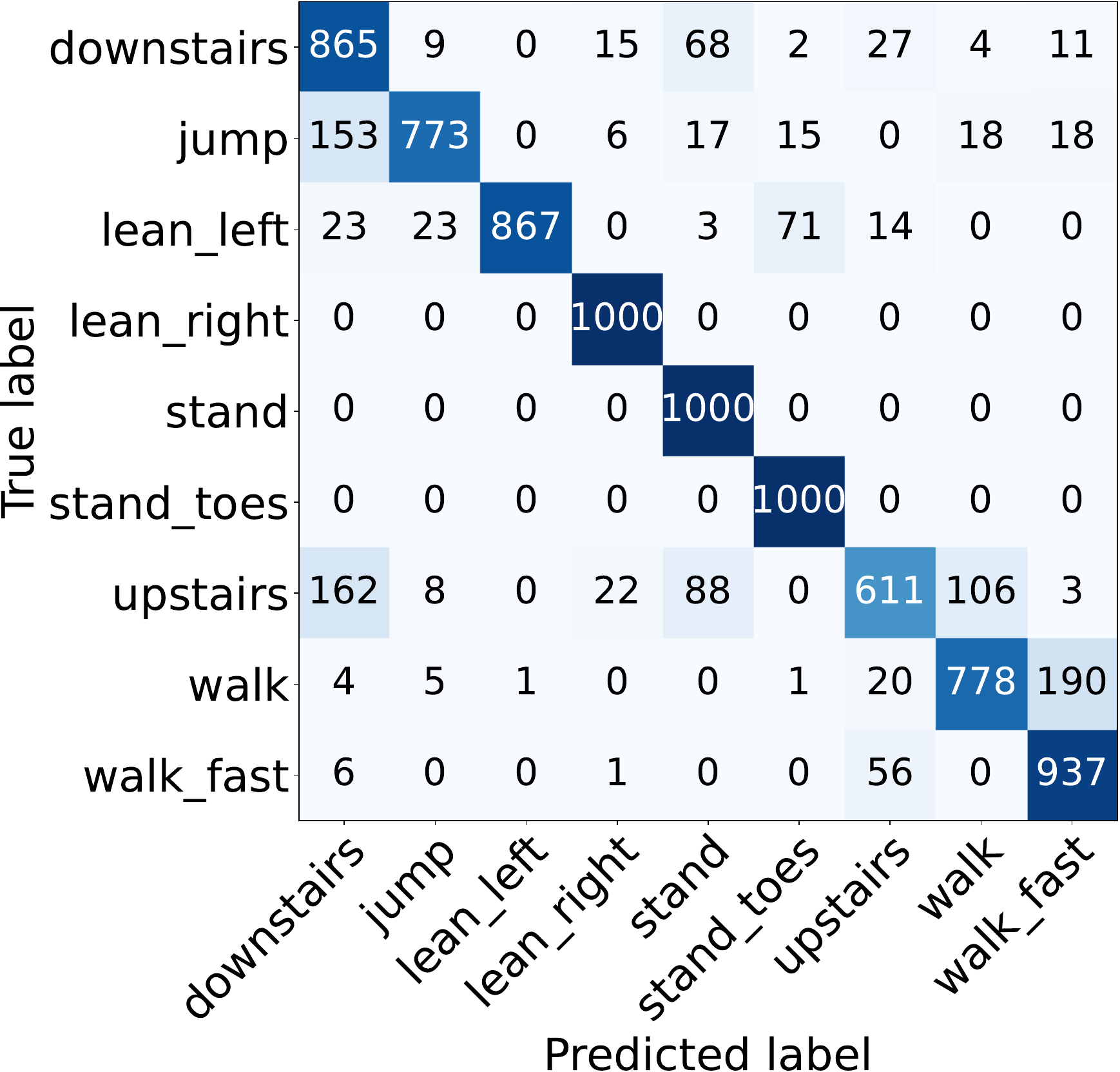}
        \label{fig:tstransformer}
    }
    \subfigure[VideoMAE]{
        \includegraphics[width=4cm]{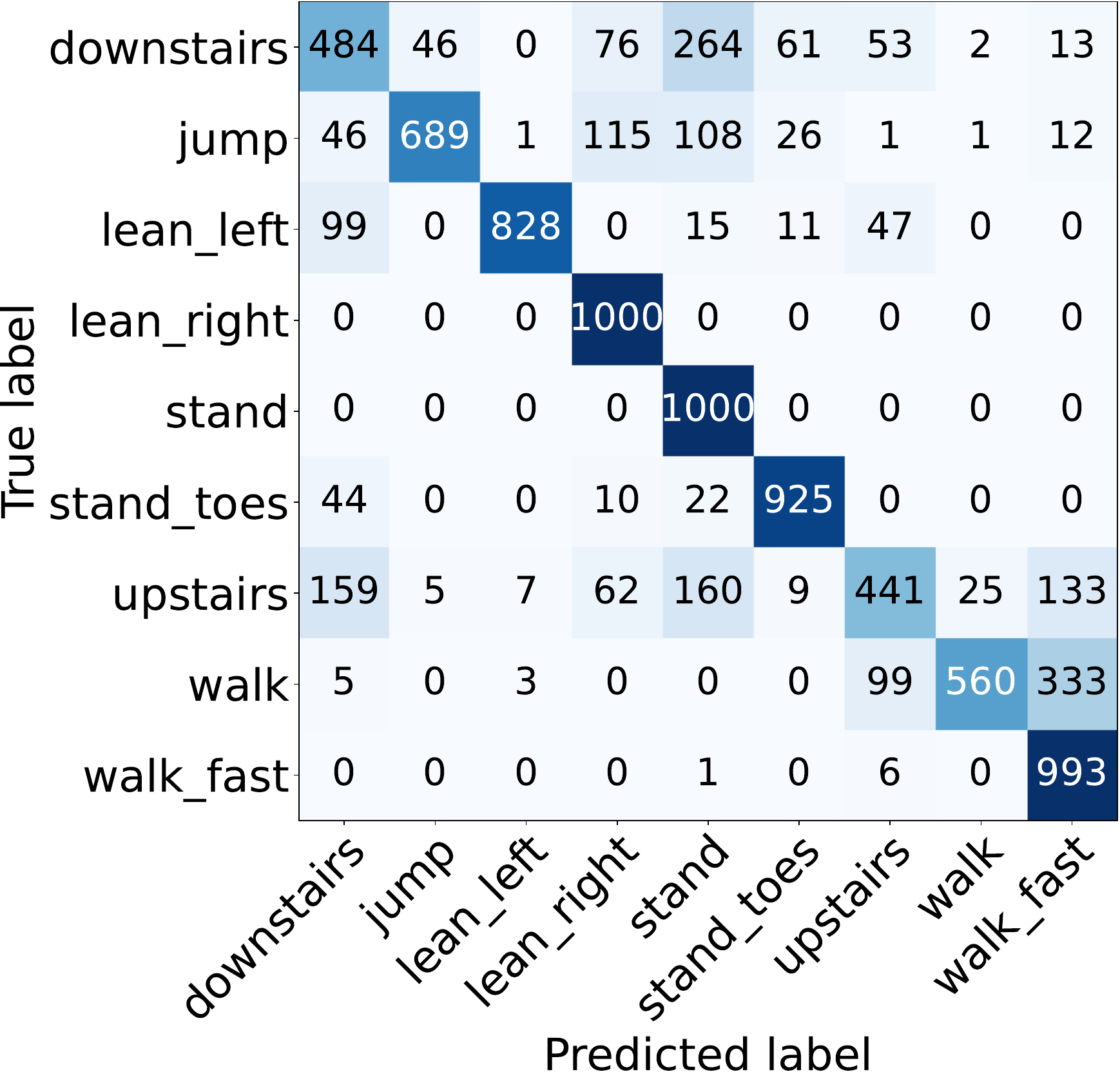}
        \label{fig:ours}
    }
    \subfigure[STAT]{
        \includegraphics[width=4cm]{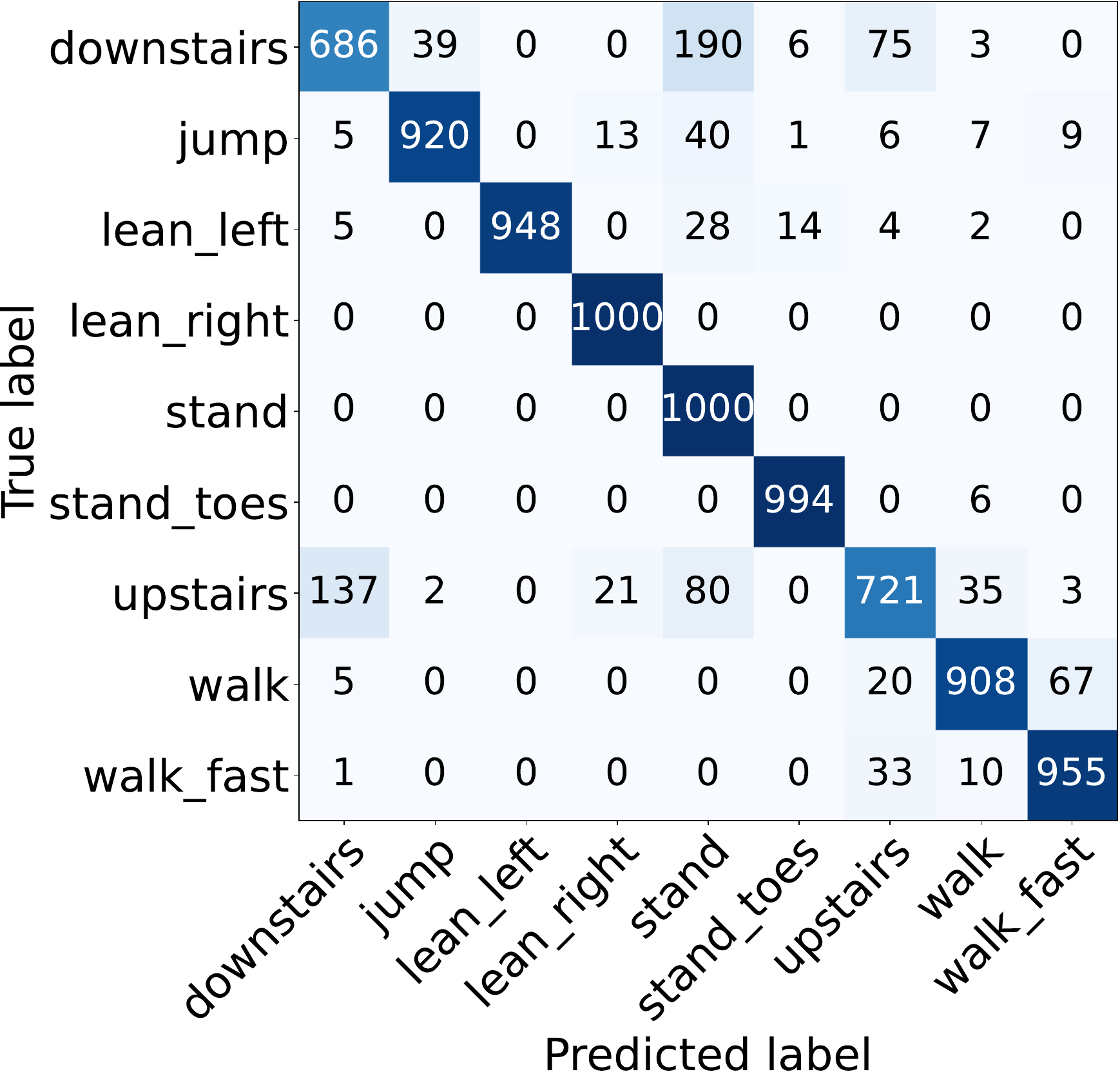}
        \label{fig:statransformer}
    }
    \caption{Confusion Matrices of all models.}
    \label{confusion matrix}
\end{figure*}

\section{Experiments}
\subsection{Experimental Settings} 
\subsubsection{Dataset}

As tactile action classification is a promising new application scenario that is under development, there is only one large-scale open dataset by far. So our experiments are conducted on the public tactile signal dataset\footnote{http://senstextile.csail.mit.edu/}, which is collected by individuals with two wearable electronic socks to perform specific actions.  The dataset consists of tactile signals with 9 labeled actions, namely walking, leaning on the left foot, leaning on the right foot, climbing downstairs, climbing upstairs, jumping, standing on toes, fast walking, and standing upright. 
The statistics are shown in Table \ref{dataset}. $T$, $H$, and $W$ are set to 45, 32, and 32, respectively.

As the sampling frequency is 15HZ, each piece of data is collected in 3 seconds. Following the providers' settings, 500 and 1,000 samples of each action are used in validation and testing, respectively, and the other samples are used in training (each action type will be sampled to 4,000 samples). Only the training set will be adopted for model pretraining to avoid data leakage. 

\subsubsection{Metrics}
We use accuracy and macro-F1 as evaluation metrics. As there are multiple classes, we report both Top-1 \& Top-3 accuracy as in previous studies~\cite{luo2021learning}. Besides, we add Macro-F1 to show the comprehensive performance on the imbalanced dataset of all models.

\subsubsection{Baselines}
To demonstrate the effectiveness of our model, we use several state-of-the-art baselines: 
\begin{itemize}
    \item CNN\&GRU~\cite{luo2021learning}: This method adopts convolution and recurrent networks for action classification with tactile signals; 
    \item {TST}~\cite{DBLP:conf/kdd/ZerveasJPBE21}: TST is a state-of-the-art transformer-based model for continuous multivariate signal classification with pretraining;
    \item {VideoMAE}~\cite{DBLP:journals/corr/abs-2203-12602}: This is a state-of-the-art video classification model with masked auto-encoders. 
\end{itemize}

\subsubsection{Implementation Details}

\begin{table}[]
\centering
\small
\begin{tabular}{l|r}
\toprule
\textbf{Hyper-parameters} & \textbf{Values} \\ \midrule
\#Comparison Pairs $N_{\mathrm{comp}}$       & 10, 20, 30, 40, 50     \\
Loss Weight  $\beta$        & 0.5, 0.75, 1, 1.5, 2, 2.5  \\
Masking Ratio       & 0.1 - 0.9 with step length 0.1 \\
Adam Learning Rate  & 1e-3, 5e-3, 1e-2             \\
Transformer Layer $K$       & 3, 6, 9, 12            \\
 \bottomrule
\end{tabular}%
\caption{Summarization of tuned hyper-parameters.}
\label{hyperparameter}
\end{table}
We tune hyper-parameters as shown in Table \ref{hyperparameter}. In addition, the tubelet parameters $L$ and $P$ are set to 5 and 4, while the pretraining and fine-tuning epoch is set to 60. The embedding dimension $D$ is set to 768, in which batch size is 64 and weight decay is 1e-4.
For baselines, we employ their public implementations and tune them with hyperparameters suggested by their authors. 

All experiments are implemented by Pytorch 1.7 and executed on 4 Tesla V100 or GeForce RTX 3090 GPUs. Note that only the training data is used for the pretraining of TST, VideoMAE, and STAT to avoid data leakage. Experiments are repeated 5 times with different random seeds. Besides, the total training time of STAT is similar to VideoMAE (10 hours). The code is available at https://github.com/Aressfull/sock\_classification.

\subsection{Overall Performances}

Experimental results of our STAT and baselines are reported in Table \ref{main_result}. TST, VideoMAE, and STAT models are pretrained with the training set, and STAT w/o pretraining is directly trained for the classification task.

Firstly, our pretrained STAT outperforms all baseline models in all metrics, showing that jointly modeling spatial and temporal features contributes to better action classification results. 
STAT achieves 2.7\%, 2.0\%, and 3.1\% improvements than the best baseline in ACC@1, ACC@3, and Macro-F1, respectively. 
Secondly, STAT without pretraining performs worse than most baselines, showing that our pretraining provides significant improvements for STAT in the action classification task. 
Thirdly, for the baseline models, the widely used tactile CNN\&GRU model achieves comparable results as TST, showing that modeling spatial features and temporal features are both important in action classification. 
However, VideoMAE model performs the worst, which indicates that simply reusing the video transformer will get worse performance. The reason should be videoMAE cannot capture the translation variant of tactile signals, and tactile signals are more dense than videos (VideoMAE only uses 8 frames but uses 45 frames here).

To further analyze the performances of different models in various classes, we show the confusion matrices of all models on the test set in Figure~\ref{confusion matrix}. 
From the figures, we have the following observations:
Firstly, CNN\&GRU performs worse in temporal and spatial sensitive classes, i.e., upstairs and lean left, showing the weaknesses of current tactile classification models. Specifically, CNN\&GRU classifies many upstairs samples as walk fast due to their similar spatial features (as shown in Figure \ref{fig:compare}(b)), while our STAT can distinguish these actions more accurately due to the modeling of temporal features.
Secondly, TST performs even worse than CNN\&GRU in many actions, indicating that focusing on modeling temporal features is not enough for tactile signals. For example, TST mistakes a number of lean left samples as stand on toes because they have similar temporal features (as shown in Figure \ref{fig:compare}(a)). Our STAT rarely makes mistakes on these actions as they are distinct in spatial features.
Thirdly, our STAT model performs the best in most classes, as we jointly capture both the spatial and temporal properties of tactile signals. Meanwhile, due to the translation variant property of tactile signals, VideoMAE, which is designed for video classifications, is unsuitable for this task.

\subsection{Analyses}
\subsubsection{Ablation Study}

To verify the effectiveness of the designed pretraining task and embeddings, we conduct ablation studies. Table~\ref{tab:ablation} shows our ablation strategies and their performances. Note that the MTR pretraining task, position, and tubelet embeddings are used in all experiments, as we focus on analyzing the newly designed models here.

We have the following observations from the results: 
Firstly, all designed modules contribute to the classification task, as 
STAT (Strategy 5) achieves the best performance with all modules in ACC@1 \& Macro-F1, and comparable results in ACC@3.
Secondly, by comparing Strategies 1,2,3 in pairs, we find that removing any one of the two designed embeddings will result in a large drop in performance. Besides, temporal embeddings are more important than spatial embeddings, as Strategy 1 performs better. 
Thirdly, STAT with both embeddings (Strategy 3) outperforms STAT with only the temporal task (Strategy 4) in all metrics. This indicates that only adopting the proposed pretraining task cannot make full use of its ability.

\begin{table}[]
\centering
\resizebox{\columnwidth}{!}
{%
\begin{tabular}{c|ccc|rrr}
\toprule
\# & \textbf{\begin{tabular}[c]{@{}c@{}}TE\end{tabular}} & \textbf{\begin{tabular}[c]{@{}c@{}}SE\end{tabular}} & \textbf{\begin{tabular}[c]{@{}c@{}}TPT\end{tabular}} & \textbf{ACC@1}  & \textbf{ACC@3}  & \textbf{Macro-F1} \\ 
\midrule
1  & \checkmark  &    &   & 0.8764    & 0.9678    & 0.8715    \\
2  &    & \checkmark  &   & 0.8417    & 0.9506    & 0.8337    \\
3  & \checkmark  & \checkmark  &    & 0.8947   & \textbf{0.9854}   & 0.8957 \\
4  &   &   & \checkmark   & 0.8299    & 0.9507    & 0.8247          \\
\multicolumn{1}{l|}{5} & \checkmark                                                                     & \checkmark                                                                    & \checkmark                                                                & \textbf{0.9033} & 0.9830 & \textbf{0.9015}   \\ \bottomrule
\end{tabular}%
}
\caption{Experimental results of various ablation strategies. TE: Temporal Embeddings, SE: Spatial Embeddings, and TPT: Temporal Pretraining Task.}
\label{tab:ablation}
\end{table}

\subsubsection{Hyper-parameter Analyses}
Due to the space limit, we only show two conducted hyper-parameter experiments.

\begin{figure}[t]
    \centering
    \includegraphics[width=0.7\linewidth]{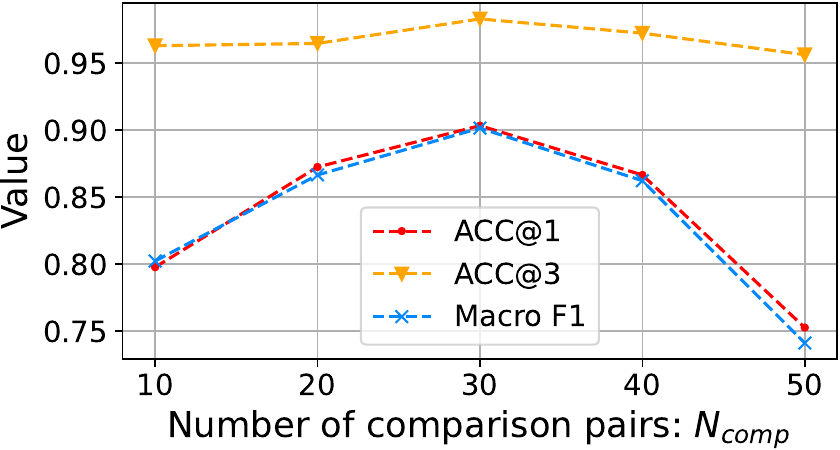} 
    \caption{Effect of the number of comparison pairs.}
    \label{fig:comparisons}
\end{figure}
\begin{figure}[t]
    \centering
    \includegraphics[width=0.7\linewidth]{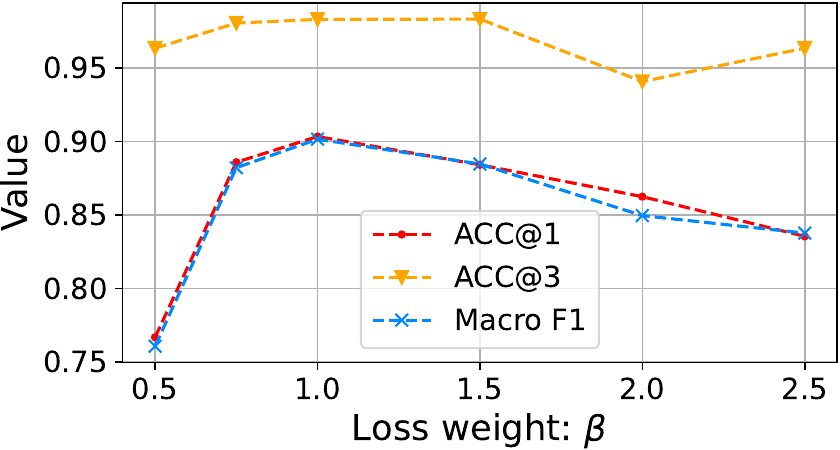} 
    \caption{Effect of the weight of temporal pretraining loss.}
    \label{fig:beta}
\end{figure}

\textbf{Effect of the Number of Comparison Pairs $N_{\mathrm{comp}}$.}
To verify the effect of $N_{\mathrm{comp}}$ in the temporal pretraining task, we conduct analyses experiments and summarize the results in Figure~\ref{fig:comparisons}.
The best performance is achieved when $N_{\mathrm{comp}}=30$. Fewer comparison pairs perform worse may be caused by insufficient training, while more pairs will not contribute to better results either.

\textbf{Effect of the Loss Weight $\beta$.}
We adjust the weight $\beta$ for our temporal pretraining task in Equation~\eqref{loss} in different values, and the results are shown in Figure~\ref{fig:beta}. It indicates that a too-low or too-high value of $\beta$ will hurt the performance of our STAT model, and $\beta=1$ performs the best.

\section{Conclusions}

Tactile signals are essential in modeling and understanding user behavior in various scenarios. 
However, neither previous tactile classification models nor transformer models for other continuous signals fail to simultaneously capture the spatial and temporal features of tactile signals. 
In this study, we propose a spatio-temporal aware tactility transformer to jointly model both spatial and temporal properties of tactile signals for action classification tasks. Spatial and temporal embeddings are designed to capture the translation variance and sequential features, respectively. Additionally, the proposed temporal pretraining task explicitly models the time order features. Experimental results show that our model outperforms all baseline models in all metrics. 

Our model shows promising performance and can contribute to better utilizing tactile signals in other scenarios. In the future, we plan to introduce side information about tactile signals to achieve better performance.

\section{Ethics Statements}
This work uses public datasets and does not directly involve ethical issues. However, if the method is to be applied in scenarios of personal-related tasks, it should obtain user consent to avoid ethical problems.

\section*{Acknowledgements}
This work is supported by the National Key R\&D Program of China (2022ZD0160502) and the National Natural Science Foundation of China (No. 62002191). We appreciate all the reviewers for their insightful suggestions.

\bibliography{aaai24}

\end{document}